\PassOptionsToPackage{prologue,table}{xcolor}
\documentclass[conference,a4paper]{APSIPA2021}
\usepackage{amsmath}
\usepackage{float}
\usepackage{booktabs,caption}
\usepackage{graphicx}
\usepackage{booktabs}
\usepackage{multirow}
\usepackage[table]{xcolor}
\usepackage{hyperref}

\usepackage{threeparttable}
\usepackage[backend=biber,style=ieee,]{biblatex}
\usepackage{geometry}
\usepackage{fancyhdr}

\fancypagestyle{firststyle}{
  \fancyhf{}
  \fancyhead[C]{2024 Asia Pacific Signal and Information Processing Association Annual Summit and Conference (APSIPA ASC)}
}

\newcommand\blfootnote[1]{%
\begingroup
\renewcommand\thefootnote{}\footnote{#1}%
\addtocounter{footnote}{-1}%
\endgroup
}

\begin{document}

\title{Fine-Grained Quantitative Emotion Editing for Speech Generation}

\author{
\authorblockN{
Sho Inoue\authorrefmark{1}\authorrefmark{2},
Kun Zhou\authorrefmark{3},
Shuai Wang\authorrefmark{2}\authorrefmark{4}, and
Haizhou Li\authorrefmark{1}\authorrefmark{2},
}
\authorblockA{
\authorrefmark{1}\textit{School of Data Science} \authorrefmark{2}\textit{Shenzhen Research Institute of Big Data}\\
\textit{The Chinese University of Hong Kong, Shenzhen (CUHK-Shenzhen), China}\\
}
\authorblockA{
\authorrefmark{3}\textit{Alibaba Group, Singapore}\\
}
\authorblockA{
E-mail: shoinoue@link.cuhk.edu.cn, kun.z@alibaba-inc.com, wangshuai@cuhk.edu.cn, and haizhouli@cuhk.edu.cn
}

}

\maketitle
\blfootnote{
\authorrefmark{4}Corresponding Author\\
The research is supported by National Natural Science Foundation of China (Grant No. 62271432); Internal Project of Shenzhen Research Institute of Big Data (Grant No. T00120220002); Shenzhen Science and Technology Program ZDSYS20230626091302006; Shenzhen Science and Technology Research Fund (Fundamental Research Key Project Grant No. JCYJ20220818103001002); and CCF-NetEase ThunderFire Innovation Research Funding (No. CCF-Netease 202302).
}

\thispagestyle{firststyle}
\pagestyle{fancy}
\begin{abstract}
It remains a significant challenge how to quantitatively control the expressiveness of speech emotion in speech generation. In this work, we propose an approach for quantitative manipulation of the emotion rendering for emotion editing in speech generation. We apply a hierarchical emotion distribution extractor, i.e. Hierarchical ED, that quantifies the intensity of emotions at different levels of granularity. Hierarchical ED is subsequently integrated into the FastSpeech2 framework, guiding the model to learn emotion intensity at phoneme, word, and utterance levels. During synthesis, users can manually edit the emotional intensity of the generated voices. Both objective and subjective evaluations demonstrate the effectiveness of the proposed network in terms of fine-grained quantitative emotion editing. 
\end{abstract}

\begin{figure*}
\centerline{\includegraphics[width=17cm]{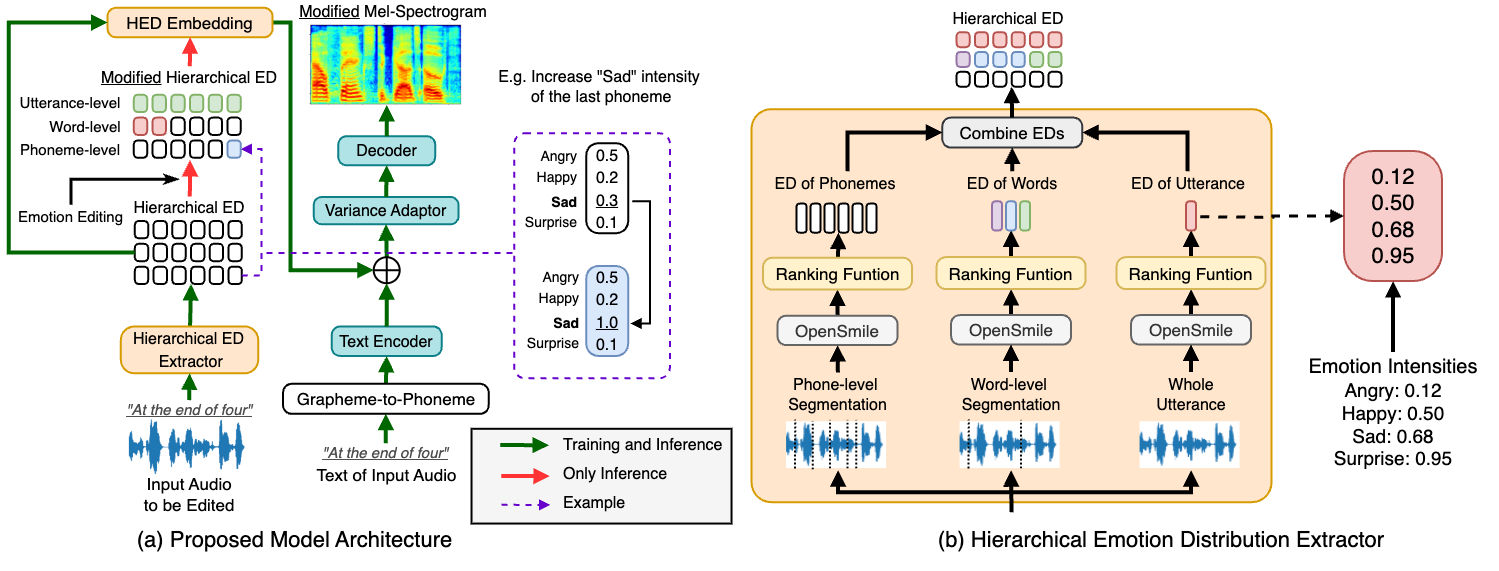}}
\caption{(a) Model architecture with hierarchical emotion distribution (HED) mechanism in TTS. During inference, the framework extracts hierarchical ED from input audio (``emotion editing''). Users can manually modify hierarchical ED to control emotion intensities at phoneme, word, and utterance levels; 
(b) Hierarchical ED extraction workflow for emotion distributions at phoneme, word, and utterance levels. 
}
\label{fig:training}
\vskip -1em
\end{figure*}

\section{Introduction}
\IEEEPARstart{S}{peech} emotion, characterized by prosodic patterns at phoneme, word, and utterance level~\cite{triantafyllopoulos2023overview, kun2022emotion, hirschberg2006pragmatics}, is effectively represented through a hierarchical structure \cite{el2011survey,schuller2018speech}. These patterns, including pitch, tempo, rhythm, stress, lexical emphasis, and speaker individuality, are crucial for accurate emotion rendering~\cite{hieprosody, zhou2020converting,zhou2021vaw,du2021disentanglement}.
Existing methods, like style tokens, represent prosody patterns but often lack interpretability, quantitativeness, and fine-grained controllability~\cite{plpm, tits2022controlling}, highlighting the need for more effective models.
In neural text-to-speech, speech editing techniques empower models to alter audio segments based on user instructions, enabling fine-grained prosody prediction in synthesized speech~\cite{editts,contextaware}. However, this approach is deficient in quantitative control.

The complexity of speech emotions, as evidenced by various studies~\cite{zhou2023mixedevc, xu2011speech, latorre2008multilevel, zhou2020transforming}, indicates that merely adjusting prosodic features is inadequate.
Additionally, the hierarchical nature of speech emotion~\cite{hieprosody}, with unique characteristics at utterance~\cite{hy-utt}, word~\cite{hy-word2}, and phoneme~\cite{hieprosody} levels, calls for a multi-level approach for accurate and human-like emotional expression in speech synthesis. Previous studies on text-to-speech and emotional voice conversion~\cite{ming2016deep,zhou2020transforming,lei2022multiscale} also highlight the importance of multi-level modeling for speech emotions.
Thus, a quantifiable method capable of modeling the hierarchical structure of emotions is imperative for effective and nuanced emotion rendering in speech synthesis.

%In this study, we present a novel approach for fine-grained emotion editing, offering an intuitive and quantitative mechanism to manipulate emotion rendering at phoneme, word, and utterance levels. 

%In our previous study \cite{ShoICASSP}, we have investigated a fine-grained emotion controlling approach for text-to-speech (TTS) systems. We further extend the approach in this paper to facilitate quantitative emotion editing, which provides an intuitive and quantitative way to manipulate emotion rendering at the phoneme, word, and utterance levels.
%In this study, we extend a novel approach for fine-grained, emotion-controllable text-to-speech (TTS)\cite{ShoICASSP} to facilitate emotion editing. 
%This method provides an intuitive and quantitative way to manipulate emotion rendering at the phoneme, word, and utterance levels. 
%Compared to our previous work~\cite{ShoICASSP}, which only integrates fine-grained embedding into the variance adaptor of FastSpeech2, 
%limiting its applicability to other TTS frameworks, 
%our emotion editing approach could enable fine-grained and flexible human editing to any TTS model.
In our previous study~\cite{ShoICASSP}, we investigated a fine-grained emotion control approach for text-to-speech (TTS) systems. In this paper, we further extend the approach to enable quantitative emotion editing. This provides an intuitive and quantitative method for manipulating emotion rendering at the phoneme, word, and utterance levels. Compared to our previous work~\cite{ShoICASSP}, which integrated fine-grained embedding solely into the variance adaptor of FastSpeech2, our proposed emotion editing approach allows for fine-grained and flexible human editing across any TTS model.

Our contributions are summarized as follows:
\vskip-0.8em
{
\setlength{\leftmargini}{15pt} 
\begin{itemize}
\setlength{\itemsep}{5pt}      %2. ブロック間の余白
\setlength{\parskip}{-5pt}      %4. 段落間余白．
\setlength{\itemindent}{0pt}   %5. 最初のインデント
\setlength{\labelsep}{5pt}     %6. item と文字の間
\item We propose a novel fine-grained emotion editing framework that facilitates the emotion rendering in previously unseen audios and allows for the adjustment of emotion intensity for a speech generation framework;

\item We thoroughly compare our approach with the baseline method of phoneme-level emotion intensity control, demonstrating our model's superior emotion expressiveness and controllability.

%\item We devise a Hierarchical Emotion Distribution extractor that hierarchically models emotion intensity, effectively capturing both fine-grained and global emotional variations. This extractor automatically predicts emotion distributions across different levels of granularity;

\item At run-time, users have the flexibility to analyze and manipulate the emotion distribution from the audio sample (``emotion editing''). This approach empowers users with a quantitative method to modify emotion rendering of a spoken utterance.
\end{itemize}
}

The rest of this paper is organized as follows: In Section 2, we discuss the related works. Section 3 describes our proposed methodology. In Section 4, we report our experiments and results. Section 5 concludes our study.

\section{Related Works}

\subsection{Control of Speech Emotion}
There has been a growing interest in enabling the control of synthesized emotions~\cite{emogst,emoscaler,PADcontrol}.
Emotion-enhanced GST~\cite{emogst} incorporated an emotion recognition task to facilitate the modeling of emotion-related prosody. Some studies explored quantitative methods for controlling emotions through relative attributes~\cite{9003829,kun-intensity, kun-mix, pmsemotts}. Another study~\cite{interintra} delved into inter- and intra-class distances to achieve fine-grained control over recognizable intensity differences. EmoQ-TTS \cite{emoq} adopted a distance-based intensity quantization approach to capture emotion intensity.

Unlike MsEmoTTS~\cite{msemotts}, which covers only utterance-level emotional change with fine-grained intensity changes, this work aims to offer quantitative and hierarchical control over emotions at various levels of speech units, e.g. phoneme, word, and utterance. It provides a versatile tool for editing emotion rendering in any speech segment.

\subsection{Speech Editing}
Speech editing, particularly semantic editing, focuses on altering the textual information of speech while preserving the naturalness of the synthesized output~\cite{editts,contextaware,editspeech}. EditSpeech~\cite{editspeech} employed partial inference and bidirectional fusion techniques to effectively modify speech semantics. Another research, EdiTTS~\cite{editts}, utilized perturbations in the Gaussian space to allow for detailed audio edits, adjusting context and pitch, and enabling user-guided prosody editing. Yet another study~\cite{contextaware} presented a method for phoneme-level pitch adjustment and time-stretching, providing users with greater control over prosody.

This work is motivated to provide a systematic approach for precisely controlling high-level prosodic patterns, which differs from the prior studies which primarily focused on modifying the physical speech attributes, e.g. pitch and duration.

In this section, we delve into the details of the proposed fine-grained emotion editing mechanism. 
We begin by formulating the problems and introducing a quantitative and hierarchical emotion control mechanism. We then detail the design of the hierarchical emotion distribution (ED) extractor, along with its associated training scheme. Lastly, we explain how to render the desired emotions by editing. 

\section{Fine-Grained Quantitative Control} \label{sec:methodology}

\subsection{Problem Formulation}
Given an audio input and its text transcript, we would like to develop a model to control the emotion intensity of the speech segments, as illustrated in Fig.\ref{fig:training}(a). This approach is expected to work for the rendering of a single emotion or a mixture of emotions and to work seamlessly with various text-to-speech frameworks.
Typically, we label emotions at the utterance level in most speech databases without paying attention to the nuanced variations in emotion intensity. 
It remains a significant challenge as to how to achieve effective and quantifiable control over speech emotion. To address this, we aim to automatically generate fine-grained and quantitative intensity labels that act as `soft labels' for speech generation models. In this way, we eliminate the need for manual labeling of emotional intensity.
Our approach is effective in emotion intensity control and mixed-emotion rendering, which can be easily adapted to any speech generation frameworks including text-to-speech and voice conversion.

\subsection{Hierarchical Emotion Distribution Extractor} 

Following~\cite{ShoICASSP}, we apply a hierarchical emotion distribution extraction module consisting of an OpenSmile feature extractor and a pre-trained ranking function for each segmental level, which automatically quantifies the intensities of each emotion type in an utterance. 
This approach, grounded in the concept of relative attributes \cite{parikh2011relative}, allows us to measure the prominence of specific emotions in speech. By treating emotion style as a speech attribute, we model and rank the presence of various emotions relative to the other emotions.

In particular, we define the ranking function as $f(\mathbf{x}_i) = \mathbf{w}^T \mathbf{x}_i + b$, where $\mathbf{x}_i$, $\mathbf{w}$, and $b$ denote the acoustic features of the $i$-th training sample, the weight vector, and bias, respectively. The parameters ($\mathbf{w}$ and $b$) are optimized using the support vector machine's objective function for binary classification (e.g. Angry and Non-angry classification)~\cite{svm}.
To provide a quantifiable measure, we further normalize the values obtained from the ranking function to a range of $[0,1]$, where a larger value represents a stronger emotion intensity. 
Our methodology facilitates the labeling of training data with continuous emotion intensities across various emotions. Additionally, it allows for the quantification and regulation of emotion intensity in unseen utterances during run-time.

The hierarchical ED extractor consists of pre-trained ranking functions as illustrated in Fig.\ref{fig:training}(b).
The extractor first segments the input audio signal into phoneme, word, and utterance levels via Montreal Forced Aligner~\cite{mfa}. Then, it extracts an 88-dimensional audio feature set for each temporal segment with OpenSMILE ~\cite{opensmile}.
The pre-trained ranking function automatically estimates an ED vector for each audio segment. Each value in this vector represents the intensity of a specific emotion encoded in the audio segment. In practice, the utterance-level emotion distribution is duplicated across all phonemes, and the word-level emotion distribution is replicated across the corresponding phonemes. 
Once formulated, the hierarchical ED vectors are integrated with linguistic embeddings in the variance adaptor during training.

\subsection{Hierarchical Emotion Intensity Modeling}

We integrate the aforementioned hierarchical ED extractor into the FastSpeech2 framework, which comprises a text encoder, variance adaptor, and decoder, as depicted in Fig.\ref{fig:training}(a). 
During training, given an input pair $<$audio, text$>$, this extractor predicts the emotion intensity at phoneme, word, and utterance levels from the audio. These predictions form a hierarchical ED as illustrated in Fig.\ref{fig:training} (b). The text encoder converts the phoneme sequence into text embeddings. The variance adaptor, utilizing both text and ED embeddings, predicts pitch, duration, and energy. Subsequently, the decoder reconstructs the Mel-spectrogram, guided by an L1 loss function. This process enables the framework to learn the hierarchical nature of emotion intensity and to establish a link between linguistic and hierarchical ED information.

\subsection{Emotion Editing}

During run-time, as depicted in Fig.~\ref{fig:training}(a), our framework enables emotion editing. For unseen $<$audio, text$>$ inputs, it derives the hierarchical Emotion Distribution (ED) from the audio signal. Users can then edit the emotion rendering by adjusting emotion distributions at three different levels (Hierarchical ED).

\section{Experiments and Results}
\subsection{Experimental Setup} \label{sec:dataset}
We conduct our experiments on two datasets: Blizzard Challenge 2013 dataset, i.e. Blizzard~\cite{blizzard}, and Emotion Speech Dataset, i.e. ESD~\cite{zhou2022emotional,zhou2021seen}.
Blizzard, audiobooks read by a single female speaker, offers approximately 42 hours of expressive speech with varied prosody but lacks emotion labels. We randomly select 200 samples from four audiobooks for testing. On the other hand, ESD comprises over 29 hours of emotional speech in five categories---Neutral, Angry, Happy, Sad, and Surprise---from 20 speakers, including 10 native English and 10 Mandarin speakers. Our study exclusively employs English recordings. For training the hierarchical ED extractor's ranking functions, we select 20 random samples per speaker and emotion from ESD.

\subsection{Model Architecture}
We use FastSpeech2 as the backbone framework~\cite{fastspeech2}, which consists of a text encoder, variance adaptor, and decoder. A transformer~\cite{transformer} based text encoder converts an input phoneme sequence into a linguistic embedding. Variance adaptors use feed-forward networks for the duration, pitch, and energy prediction. A transformer-based decoder synthesizes a mel-spectrogram from the variance-adapted features. The loss function combines the L1 loss between the predicted and target mel-spectrogram and the mean squared error loss from predicted prosodic features. 
We opt for the Adam optimizer~\cite{adam}, setting a batch size of 32 and conducting 800,000 iterations of training over 48 hours on a single GPU. The ED embedding layers comprise fully connected layers with a Tanh activation function. We adopt the text-based emotion intensity predictor from MsEmoTTS~\cite{msemotts}, which uses two 1D convolution layers, layer normalization, and dropout. HiFiGAN~\cite{hifigan} serves as the vocoder, pre-trained on the Blizzard dataset.

\begin{figure*}
\centerline{\includegraphics[width=17cm]{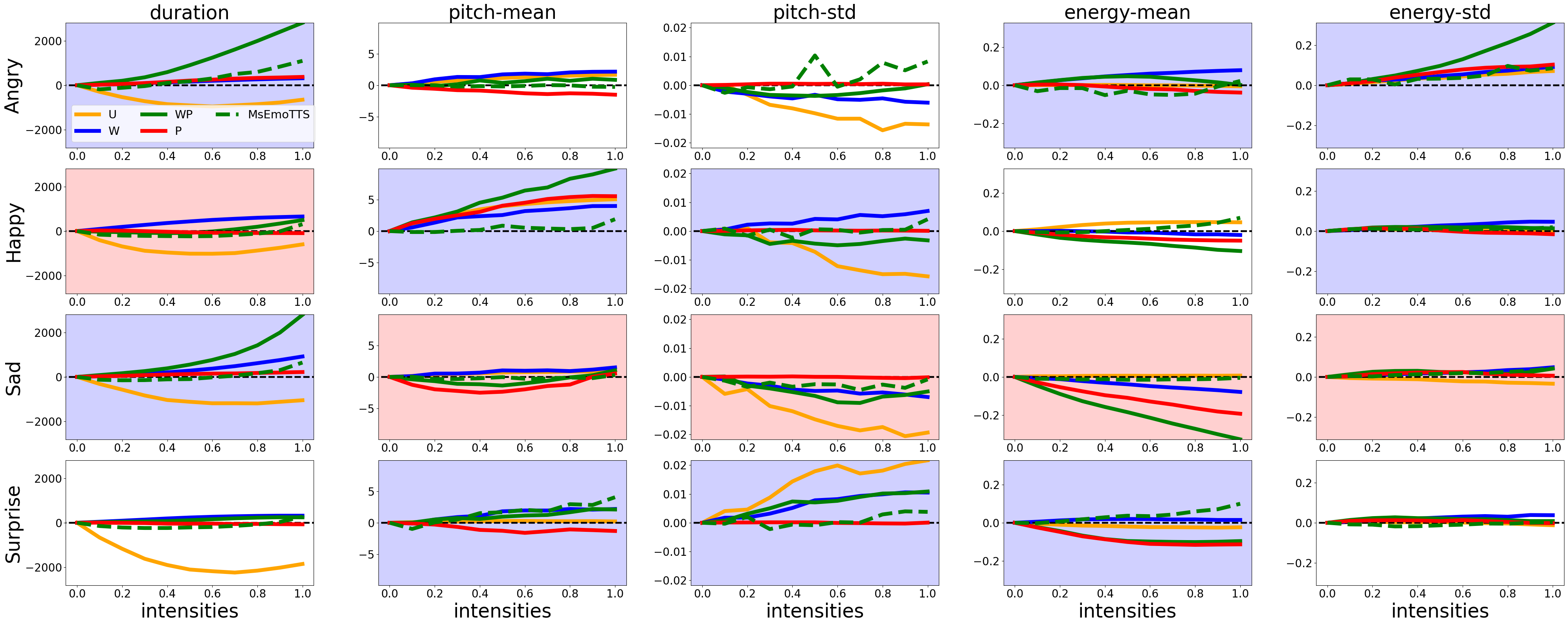}}
\caption{
The illustration of prosodic variants with intensity changes. The red background represents the expected negative trend, the blue indicates the expected positive trend, both summarized from the ESD dataset. `U', `W', `P', and `WP' indicate the controlling segments in Hierarchical ED (our model): Utterance, Word, Phoneme, and Word-and-Phoneme, respectively.
}
\label{fig:prosody_model}
\end{figure*}

\subsection{Results and Analysis}
We conduct both objective and subjective evaluations, focusing on speech quality, emotional expressiveness, and emotion controllability.
We integrate MsEmoTTS~\cite{msemotts} into the FastSpeech2 framework as the baseline.
Our subjective evaluation involves listening experiments with 18 participants, each listening to 120 synthesized samples under specific guidelines. 
We suggest readers to refer to our demo page\footnote{
\textbf{Speech Demos}: \url{https://shinshoji01.github.io/Hierarchical-ED-Demo/}
}.

\subsubsection{Speech Quality}
We conduct a Mean Opinion Score (MOS) test to evaluate the overall speech quality,
where a higher MOS represents better speech quality.
As shown in Table~\ref{table:mos}, our model demonstrates superior performance over the baseline, as indicated by its consistently higher MOS scores.

\begin{table}[t]
\caption{MOS with 95\% confidence interval}
\label{table:mos}
\begin{center}
\scalebox{1.0}{
\begin{tabular}{c|c|c}
\toprule
Ground Truth &  MsEmoTTS & \textbf{Hierarchical ED (ours)}\\
\midrule
4.211$\pm$0.142 & 3.544$\pm$0.138 & \textbf{3.596$\pm$0.141} \\
\bottomrule
\end{tabular}
}
\end{center}

\end{table}

\subsubsection{Emotion Expressiveness}
We conduct A/B preference tests where participants select the audio that more closely matches the reference in terms of emotional expressiveness. Both our model's hierarchical ED and the baseline's emotion intensity are derived from the same reference audio. As indicated in Table~\ref{table:preference}, our model outperforms the baseline with a preference rate of 43.51\%.

\begin{table}[t]
\caption{A/B preference test for emotion expressiveness with 95 \% confidence interval.}

\label{table:preference}
\begin{center}
\scalebox{1.}{
\begin{tabular}{c|c|c}
\toprule
MsEmoTTS & Neutral & \textbf{Hierarchical ED (ours)}\\
\midrule
28.77$\pm$5.256 & 27.72$\pm$5.197 & \textbf{43.51$\pm$5.756} \\
\bottomrule
\end{tabular}
}
\end{center}
\end{table}

To further objectively evaluate emotion expressiveness, we calculate various metrics: (1) Mel-Cepstral Distortion (MCD)\cite{mcd} to measure spectral similarity, (2) Pitch/Energy Distortion for assessing prosody alignment, and (3) Frame Disturbance (FD)\cite{fd} to evaluate duration deviation. The results summarized in Table~\ref{table:mcd} confirms our framework's superiority over the baseline MsEmoTTS in replicating target emotions, thereby validating its effectiveness in emotion modeling.

\begin{table}[t]
\caption{Objective evaluation for emotion expressiveness}
\label{table:mcd}
\begin{center}
\scalebox{0.88}{
\begin{tabular}{c||c|c|c|c}
\toprule
%& MCD $\downarrow$& Pitch $\downarrow$ ($*10^1$) & Energy $\downarrow$ ($*10^{-2}$) & FD $\downarrow$ ($*10^1$)\\
& MCD & Pitch  ($10^1$) & Energy  ($10^{-2}$) & FD ($10^1$)\\
\midrule
MsEmoTTS &4.783$\pm$0.135 & 1.215$\pm$0.064 & 4.884$\pm$0.225 & 7.841$\pm$0.842\\
\textbf{HED (ours)} & \textbf{4.348$\pm$0.109} & \textbf{1.151$\pm$0.052}& \textbf{4.018$\pm$0.162}& \textbf{6.881$\pm$0.710}\\
\bottomrule
\end{tabular}
}
\end{center}
\vskip -2em
\end{table}

\subsubsection{Emotion Controllability}

We conducted best-worst scaling (BWS) tests~\cite{bws} to compare word-level emotion controllability between our model and the baseline. In these tests, we modified the emotion intensity of three words in each utterance to 0.0, 0.5, and 1.0. Evaluators listened to the audio samples and selected the least and most expressive ones. As shown in Table~\ref{table:control}, our model demonstrated a distinct preference for the least expressive sample at the lowest intensity and the most expressive sample at the highest intensity, particularly for Sad and Surprise emotions. This result highlights our model's capability to effectively differentiate intensity variations.

\begin{table}[t]
\caption{
BWS Test Result: The value represents evaluator preferences (\%), with red and blue indicating the heatmap for the least expressive and most expressive audio, respectively.
}
\label{table:control}
\begin{center}
\scalebox{1.0}{
\begin{tabular}{m{0.3cm}c||m{0.4cm}m{0.4cm}m{0.4cm}m{0.4cm}|m{0.4cm}m{0.4cm}m{0.4cm}m{0.4cm}|}
& & \multicolumn{4}{c|}{Hierarchical ED (ours)} & \multicolumn{4}{c|}{MsEmoTTS}\\
& & Ang & Hap & Sad & Sur & Ang & Hap & Sad & Sur \\
\midrule
 & 0.0 & \cellcolor{red!91}{79} & \cellcolor{red!73}{63} & \cellcolor{red!77}{67} & \cellcolor{red!94}{81} & \cellcolor{red!48}{42} & \cellcolor{red!37}{32} & \cellcolor{red!24}{21} & \cellcolor{red!38}{33} \\
Least & 0.5 & \cellcolor{red!0}{0} & \cellcolor{red!32}{28} & \cellcolor{red!16}{14} & \cellcolor{red!16}{14} & \cellcolor{red!54}{47} & \cellcolor{red!54}{47} & \cellcolor{red!62}{54} & \cellcolor{red!48}{42} \\
 & 1.0 & \cellcolor{red!24}{21} & \cellcolor{red!10}{9} & \cellcolor{red!22}{19} & \cellcolor{red!5}{5} & \cellcolor{red!12}{11} & \cellcolor{red!24}{21} & \cellcolor{red!29}{25} & \cellcolor{red!29}{25} \\
\midrule
 & 0.0 & \cellcolor{cyan!12}{11} & \cellcolor{cyan!20}{18} & \cellcolor{cyan!18}{16} & \cellcolor{cyan!8}{7} & \cellcolor{cyan!12}{11} & \cellcolor{cyan!13}{12} & \cellcolor{cyan!34}{30} & \cellcolor{cyan!20}{18} \\
Most & 0.5 & \cellcolor{cyan!18}{16} & \cellcolor{cyan!8}{7} & \cellcolor{cyan!10}{9} & \cellcolor{cyan!8}{7} & \cellcolor{cyan!30}{26} & \cellcolor{cyan!18}{16} & \cellcolor{cyan!26}{23} & \cellcolor{cyan!34}{30} \\
 & 1.0 & \cellcolor{cyan!86}{74} & \cellcolor{cyan!87}{75} & \cellcolor{cyan!87}{75} & \cellcolor{cyan!100}{86} & \cellcolor{cyan!73}{63} & \cellcolor{cyan!83}{72} & \cellcolor{cyan!54}{47} & \cellcolor{cyan!61}{53} \\
\bottomrule
\end{tabular}
}
\end{center}
\vskip -2em
\end{table}

We further validated the controllability of our model across fine-grained emotional variances, examining them at utterance, word, phoneme, and word-and-phoneme combination levels.
We increased the emotion intensity values from 0.0 to 1.0 and subsequently computed various prosody features, such as duration, mean/standard deviation of pitch, and mean/standard deviation of energy, shown in Fig.\ref{fig:prosody_model}. 
These prosodic characteristics are known to correlate significantly with emotion intensity, as indicated in the literature \cite{schuller2018speech}
For example, sadness often manifests in a slower speaking rate and lower values for pitch and energy mean/standard deviation.
We performed an analysis of ESD to explore the relationship between acoustic features and various emotions, using different colors in Figure \ref{fig:prosody_model} to indicate positive and negative correlations. 
A red background represents an expected negative trend, while a blue background signifies a positive trend as intensity increases. 
We have observed that our proposed model closely aligns with the expected trends. For example, ESD reveals a positive correlation between happiness and mean pitch, as well as a negative correlation between sadness and mean energy. Our models can effectively adjust these features as emotion intensity varies.

Furthermore, we observe significant prosodic changes when editing both word and phoneme-level emotions and the changes in the standard deviation of pitch at the utterance level align with our expectations.
The figure also illustrates that Hierarchical ED outperforms the baseline model, especially when controlling word and phoneme levels simultaneously. Our model exhibits a significantly closer alignment with the expected trend.

\section{Conclusion}

We introduce a fine-grained emotion editing approach for speech generation tasks. 
%We design a novel hierarchical ED extractor that learns emotion intensity at different granularity levels. 
At run-time, users can quantitatively control emotion rendering at phoneme, word, and utterance levels. Both objective and subjective evaluations validate the effectiveness of our proposed idea. As a future work, we aim to enhance paragraph-level information processing and develop emotional text-to-speech models featuring robust emotion consistency. 
Additionally, the Hierarchical ED framework holds potential for application in diverse speech generation frameworks and contexts, including voice conversion tasks.

%\section*{Acknowledgment}
%The research is supported by National Natural Science Foundation of China (Grant No. 62271432); Internal Project of Shenzhen Research Institute of Big Data (Grant No. T00120220002); Shenzhen Science and Technology Program ZDSYS20230626091302006; Shenzhen Science and Technology Research Fund (Fundamental Research Key Project Grant No. JCYJ20220818103001002); and CCF-NetEase ThunderFire Innovation Research Funding(No. CCF-Netease 202302).

\printbibliography
\end{document}